\documentclass[twocolumn]{revtex4}
\usepackage{amsfonts}
\usepackage{amsmath}
\usepackage{amssymb}
\usepackage{graphicx}
\usepackage{color}

\begin{document}

\title{Navier-Stokes Equation by Stochastic Variational Method}
\author{T.~Koide}
\author{T.~Kodama}
\affiliation{Instituto de F\'{\i}sica, Universidade Federal do Rio de Janeiro, C.P.
68528, 21941-972, Rio de Janeiro, Brazil}

\begin{abstract}
We show for the first time that the stochastic variational method can
naturally derive the Navier-Stokes equation starting from the action of
ideal fluid. In the frame work of the stochastic variational method, the
dynamical variables are extended to stochastic quantities. 
Then the effect of dissipation is realized as the direct consequence 
of the fluctuation-dissipation theorem. 
The present
result reveals the potential availability of this approach to describe more general 
dissipative processes.
\end{abstract}

\maketitle

\section{Introduction}

In almost all branches in physics, the variational principle is one of the
important guiding principles, and also serves as a powerful tool in practice 
\cite{lancz}. In particular, this principle is indispensable method to deal
with the symmetry property of a system, for example, the relativistic
covariance. Once the symmetry of a system is expressed as the invariance of
the action under the transformations of the symmetry in question, then the
resultant equations of motion automatically satisfies this symmetry.

The usual variational principle, however, is not applicable when
irreversible dynamics is present. Because dissipation involves energy
exchange processes between macroscopic and microscopic motions, associated
with the entropy production mechanism. These microscopic degrees of freedom
are not included as the system degrees of freedom in the Lagrangian. Thus,
dissipation is beyond the scope of the classical variational method \cite%
{lancz}. To incorporate dissipative effects, we may introduce, for example,
the Rayleigh dissipation function or some time-dependent external factor in
actions, but it is not easy to specify uniquely them \cite{becker}.

Three decades ago, a new variational approach called the stochastic
variational method (SVM) was introduced \cite{yasue1,morato}.
There, dynamical variables are extended to stochastic ones. In this way,
dissipation is naturally induced by noise, without requiring additional
functions. For example, it was shown that the Navier-Stokes (NS) equation
for viscous fluids can be derived for an incompressible case \cite{yasue2,dif}. 
SVM is also used to derive the Schr\"{o}dinger equation from
an action of classical dynamics. In the latter case, SVM has been discussed
in relation to Schr\"{o}dinger and Nelson's formulation of quantum mechanics \cite{Nelson}. 
If the general criteria of the applicability of SVM is established, it will
serve as a very powerful tool in practice. This is because, as we will see
later, once we know the form of Lagrangian of the reversible dynamics, the
effect of dissipation to this dynamics is introduced in a systematic manner.
In general, the construction of Lagrangians is 
relatively easier than that of evolution equations, because the form of the 
Lagrangian is constrained from the symmetry principle.

In spite of these interesting concepts and features of the method, studies
on its potential applicability to realistic problems have not been
sufficiently developed. For example, even for NS equation, 
only the case of an incompressible fluid 
has been studied. To claim the
reliability of SVM, it is necessary to show explicitly that this can
reproduce established dissipative equations.

In this letter, we show that the full NS equation, at least, can be derived
in the framework of SVM. Our action is the non-relativistic version of that
used for the variational formulation of relativistic hydrodynamics \cite%
{RelHydro}, but differs from those used in previous works of the classical
variational formulation of hydrodynamics \cite{lancz,yasue2}.

\section{Action for Ideal Fluid}

We first derive the Euler equation using the classical variational method.
The same action used here is employed for the derivation of the NS equation
in SVM.

Let us consider a fluid and divide it into small mass elements specified by
their Lagrangian coordinate $\mathbf{R}$'s. The space position of the fluid
element (the Euler coordinate) is denoted by 
\begin{equation}
\mathbf{r=r}(\mathbf{R},t).  \label{Euler}
\end{equation}%
Let $\mathbf{v=v}(\mathbf{R},t)$ be the velocity field of the fluid element.
When the time evolution is smooth, the time derivative of the Euler
coordinate associated with the fluid element $\mathbf{R}$ is given by this
velocity field (see later discussion), 
\begin{equation}
\left. \frac{\partial \mathbf{r}(\mathbf{R},t)}{\partial t}\right\vert _{%
\mathbf{R}}=\mathbf{v}(\mathbf{R},t).  \label{vel}
\end{equation}%
Let $U$ be the specific internal energy per unit mass of the fluid element.
In general, in local thermal equilibrium, $U$ depends on $(\mathbf{R},t)$
only through the specific entropy $\hat{s}$ and the mass density $\rho $.
With the energy density $\varepsilon $, we can write $U= \varepsilon /\rho $%
. The Lagrangian is then given by,%
\begin{equation}
L=\int d^{3}\mathbf{R\ }\rho _{0}\left( \frac{1}{2}\mathbf{v}^{2}(\mathbf{R}%
,t)-\frac{\varepsilon }{\rho }\right) ,  \label{L_Lagrange}
\end{equation}%
where the first and second terms represent, respectively, the kinetic energy
and the \textquotedblleft potential energy\textquotedblright\ associated
with the fluid element. Here, $\rho _{0}$ is the mass density measured in
the Lagrangian coordinate system and, by definition, does not depend on
time. Note that the same Lagrangian can be expressed with the integral over
the Euler coordinates as 
\begin{equation}
L=\int d^{3}\mathbf{r}\left( \frac{1}{2}\rho \ \mathbf{v}^{2}(\mathbf{R}%
,t)-\varepsilon \right) ,  \label{L}
\end{equation}%
because, from Eq.(\ref{Euler}), the two mass densities $\rho $ and $\rho
_{0} $ are related with the Jacobian $J=\det \left\vert \partial \mathbf{r}%
/\partial \mathbf{R}\right\vert $ through the coordinate transformation as $%
\rho =\rho _{0}/J$ \cite{ff}.

Then the action is expressed as 
\begin{equation}
I=\int_{t_{a}}^{t_{b}}dt\ \int d^{3}\mathbf{R\ }\rho _{0}\left( \frac{1}{2}%
\mathbf{v}^{2}-\frac{\varepsilon }{\rho }\right) .
\end{equation}%
In the usual derivation of the Euler equation, the specific entropy $\hat{s}$
should be kept constant for the variational procedure, representing that the
fluid is ideal. Then we consider only the variation of $\mathbf{r}$, leading
to 
\begin{equation}
\left. \frac{\partial \mathbf{v}}{\partial t}\right\vert _{\mathbf{R}}+\frac{%
1}{\rho }\nabla _{\mathbf{r}}P=0,  \label{euler1}
\end{equation}%
where $P$ is pressure defined by the thermodynamic relation under the
assumption of local thermal equilibrium, 
\begin{equation}
P=-\frac{d}{d(1/\rho )}\left( \frac{\varepsilon (\rho ,\hat{s})}{\rho }%
\right) _{\hat{s}},  \label{tdr}
\end{equation}%
and the notation $\nabla _{\mathbf{r}}$ represents the gradient with respect
to the Euler coordinates $\mathbf{r}$. In the following, the symbol $\nabla $
is used only for the gradient with respect to $\mathbf{r}$, and hence the
index is omitted. Equation (\ref{euler1}) is the Euler equation, since 
\begin{equation}
\left. \partial _{t}\mathbf{v}\left( \mathbf{R,}t\right) \right\vert _{%
\mathbf{R}}=\left[ \partial _{t}+\mathbf{v}(\mathbf{r},t)\cdot \nabla \right]
\mathbf{v}(\mathbf{r},t).  \label{euler2}
\end{equation}%
In the above derivation, we used the following relations, 
\begin{subequations}
\begin{eqnarray}
&&\sum_{l}\frac{\partial }{\partial \mathbf{R}^{l}}A^{il}=0, \\
&&\sum_{k}\frac{\partial \mathbf{r}^{k}}{\partial \mathbf{R}^{i}}%
A^{kj}=\sum_{k}\frac{\partial \mathbf{r}^{i}}{\partial \mathbf{R}^{k}}%
A^{jk}=J\delta ^{ij}, \\
&&\sum_{l}A^{il}\frac{\partial }{\partial \mathbf{R}^{l}}=J\nabla ^{i},
\label{cal_jder}
\end{eqnarray}%
\end{subequations}
where $A^{ij}=\partial J/\partial (\partial \mathbf{r}^{i}/\partial \mathbf{R%
}^{j})$.

\section{Stochastic Variational Method}

Following the spirit of SVM, we start from the same Lagrangian (\ref%
{L_Lagrange}) as the ideal case. In SVM \cite{yasue1,yasue2}, however, we
allow random fluctuations of $\mathbf{r}$ in its time evolution due to
noise, so that the time derivative of $\mathbf{r}$ as Eq. (\ref{vel}) is not
well-defined. The physical reason for the appearance of noise is the
microscopic degrees of freedom which are coarse-grained in macroscopic
scales. They act as the origin of fluctuations of the movement of fluid
elements, leading to dissipation. Thus, the time evolution of $\mathbf{r}$
is described by the following stochastic differential equation (SDE), 
\begin{equation}
d\mathbf{r}(t)=\mathbf{u}\ dt+\mathbf{B}\star d\mathbf{W}(t).~~(dt>0),
\label{eqn:sde}
\end{equation}%
where, in the right hand side, the first term $\mathbf{u}$ is an unknown
function of $(\mathbf{r},t)$ which is to be determined by the variational
procedure, and the last term is the noise term given by a Wiener process $%
\mathbf{W}(t)=\left( W_{x},W_{y},W_{z}\right) $ satisfying 
\begin{subequations}
\label{Wiener}
\begin{eqnarray}
E\left[ W_{j}(s)-W_{j}(u)\right] &=&0, \\
\hspace{-1cm}E\left[ (W_{j}(s)-W_{j}(u))(W_{k}(s)-W_{k}(u))\right] &=&\delta
^{jk}|s-u|, \\
E\left[ (W_{j}(s)-W_{j}(u))W_{k}(t)\right] &=&0,
\end{eqnarray}%
\end{subequations}
where $j,k=x,y,z$, $-\infty <t\leq s,u<\infty $ and $E\left[ \ \ \ \right] $
denotes the expectation value of stochastic processes. The symbol $\star $
is used to denote the Ito definition of a product for stochastic variables 
\cite{handbook}. The coefficient $\mathbf{B}$ is a vector composed of second
rank tensors. For the sake of simplicity, we consider it as constant. We
call Eq. (\ref{eqn:sde}) the forward SDE since it is defined only for $dt>0$.

In SVM, the forward SDE is not sufficient to complete the formulation of
stochastic variations. Because stochastic trajectories are not
differentiable, the definition of velocity is not unique. There are two
possible definitions for the velocity of a fluid element $\mathbf{R}$ at $t$%
, 
\begin{subequations}
\begin{eqnarray}
\mathbf{v}_{F} &\rightarrow &\lim_{dt\rightarrow 0+}\frac{\mathbf{r}(\mathbf{%
R},t+dt)-\mathbf{r}(\mathbf{R},t)}{dt},  \label{VB} \\
\mathbf{v}_{B} &\rightarrow &\lim_{dt\rightarrow 0-}\frac{\mathbf{r}(\mathbf{%
R},t+dt)-\mathbf{r}(\mathbf{R},t)}{dt}.
\end{eqnarray}%
\end{subequations}
When $\mathbf{r}$ is continuous and smooth, the two definitions should
coincide, $\mathbf{v}_{F}=\mathbf{v}_{B}$, as is the case of Eq. (\ref{vel}%
). However, stochastic $\mathbf{r}$ is not smooth and we should distinguish
the two evolutions defined by $\mathbf{v}_{F}$ and $\mathbf{v}_{B}$.

The backward SDE which describes the time reversed process of Eq. (\ref%
{eqn:sde}) is given by, 
\begin{equation}
d\mathbf{r}(t)=\tilde{\mathbf{u}}\ dt+\mathbf{B}\star d\tilde{\mathbf{W}}%
(t),~~(dt<0)  \label{eqn:sde_r}
\end{equation}%
where the new drift term $\tilde{\mathbf{u}}$ should be related to $\mathbf{%
u,}$ 
\begin{equation}
\mathbf{u}^{i}=\tilde{\mathbf{u}}^{i}+\sum_{j}2\nu ^{ij}\nabla ^{j}\ln \rho .
\label{consist_con}
\end{equation}
This relation is obtained from the consistency condition of the two
Fokker-Plank equations obtained from the two SDEs (\ref{eqn:sde}) and (\ref%
{eqn:sde_r}) \cite{yasue1,yasue2}. Here we introduced $\nu ^{ij}=$ $[\mathbf{%
B}\mathbf{B}^{T}]^{ij}/2$.~The noise term $\tilde{\mathbf{W}}$ is again the
Wiener process given by Eq. (\ref{Wiener}). There is no correlation between $%
\mathbf{W}$ and $\tilde{\mathbf{W}}$.

It should be emphasized that the velocities $\mathbf{u}$ and $\tilde{\mathbf{%
u}}$ are not parallel to the current of the mass density. From the
Fokker-Plank equation obtained from Eq. (\ref{eqn:sde}), the mass density
equation is uniquely given by 
\begin{equation}
\partial _{t}\rho =-\sum_{i}\partial _{i}(\rho \mathbf{u}^{i}-\sum_{j}\nu
^{ij}\partial _{j}\rho )=-\nabla \cdot (\rho \mathbf{v}_{m}).  \label{mass}
\end{equation}%
Here $\mathbf{v}_{m}$ is parallel to the mass current, and defined by, 
\begin{equation}
\mathbf{v}_{m}^{i}=(\mathbf{u}^{i}+\tilde{\mathbf{u}}^{i})/2=\mathbf{u}%
^{i}-\sum_{j}\nu ^{ij}\partial _{j}\ln \rho .  \label{mass_v}
\end{equation}%
We call $\mathbf{v}_{m}$ and $\mathbf{u}$ the mass velocity and the
diffusion velocity, respectively. For an incompressible fluid, all of three
velocities, $\mathbf{u},\tilde{\mathbf{u}}$ and $\mathbf{v}_{m}$ coincide
due to $\nabla \rho =0,$ showing the intrinsic difference between the
compressible and incompressible cases.

Following Ref.~\cite{yasue2}, the action expressed with stochastic variables
is then obtained from Eq.~(\ref{L_Lagrange}) by replacing $\mathbf{v}$ with
the mean forward derivative $D\mathbf{r}$, 
\begin{equation}
\hspace{-0.7cm}I=\int_{t_{a}}^{t_{b}}dt\ \int d^{3}\mathbf{R\ }\rho _{0}E%
\left[ \frac{1}{2}(D\mathbf{r}(\mathbf{R},t))\cdot (D\mathbf{r}(\mathbf{R}%
,t))-\frac{\varepsilon }{\rho }\right] .  \label{act}
\end{equation}%
Here the mean forward derivative $D\mathbf{r}$ is defined by 
\begin{equation}
D\mathbf{r}(t)\equiv \lim_{h\rightarrow 0+}E\left[ \frac{\mathbf{r}(t+h)-%
\mathbf{r}(t)}{h}\Big{|}\mathcal{P}_{t}\right] .  \label{Forward}
\end{equation}%
Here, $E[F(t^{\prime })|\mathcal{P}_{t}]$ denotes the conditional average of
the time sequence $\left\{ F\left( t^{\prime }\right) ,t_{a}<t^{\prime
}<t_{b}\right\} $, taking the expectation values for $t^{\prime }>t$, fixing 
$F\left( t^{\prime }\right) $ of $t\prime \leq t$ \cite{Nelson}. Thus $DF(t)$
is a stochastic variable. Note that the product of the mean forward
derivatives is independent of the choice of the discretization scheme such
as the Ito, Stratonovich-Fisk and H\"{a}nggi-Klimontovich schemes by the
definition of the mean forward derivative \cite{handbook}.

The mean backward derivative is, similarly, defined as 
\begin{equation}
\tilde{D}\mathbf{r}(t)\equiv \lim_{h\rightarrow 0+}E\left[ \frac{\mathbf{r}%
(t)-\mathbf{r}(t-h)}{h}\Big{|}\mathcal{F}_{t}\right] .  \label{Backward}
\end{equation}%
where, $E[F(t^{\prime })|\mathcal{F}_{t}]$ is the conditional average for
the past sequence, fixing the future values.

There exist several ways to express the classical kinetic term in terms of $D%
\mathbf{r}$ and $\tilde{D}\mathbf{r}$. For example, $D\mathbf{r} \cdot \tilde{%
D}\mathbf{r}$ is used in Ref. \cite{dif}. The final results for the equation
of motion depend on this choice. To obtain the NS equation, Eq. (\ref{act})
is employed.

Similarly to the Euler equation, we consider variations for $\mathbf{r}$ as
follows, 
\begin{equation}
\mathbf{r}_{\lambda }(\mathbf{R},t)=\mathbf{r}(\mathbf{R},t)+\lambda \mathbf{%
f}(\mathbf{r},t),  \label{vari}
\end{equation}%
where $\mathbf{f}(\mathbf{r},t)$ is an arbitrary function with the boundary
condition, $\mathbf{f}(\mathbf{r},t)_{t=t_{a}}=\mathbf{f}(\mathbf{r}%
,t)_{t=t_{b}}=0$ and $\lambda $ is a small parameter. Then, for example, the
variation of the mass density is calculated as 
\begin{equation}
\delta _{\lambda }\rho =\rho _{\lambda }-\rho =-\lambda \frac{\rho }{J}%
\sum_{ij}\frac{\partial J}{\partial (\partial \mathbf{r}^{i}/\partial 
\mathbf{R}^{j})}\frac{\partial \mathbf{f}^{i}}{\partial \mathbf{R}^{j}}%
+O(\lambda ^{2}).
\end{equation}

Keeping the terms up to first order in $\lambda $, $\delta _{\lambda }I$ is
given as 
\begin{eqnarray}
\delta _{\lambda }I &=&-\int_{a}^{b}dtd^{3}\mathbf{R}\sum_{i}\rho _{0} 
\notag \\
&&\hspace{-1cm}\times E\left[ \lambda \mathbf{f}^{i}\left( \tilde{D}D\mathbf{%
r}^{i}(\mathbf{R},t)+\frac{1}{\rho }\frac{\partial }{\partial \mathbf{r}^{i}}%
P\right) +\frac{T}{m}\delta _{\lambda } \hat{s}\frac{{}}{{}}\right] ,
\label{ls}
\end{eqnarray}%
where $T=\partial mU/\partial \hat{s}$ and $m$ is the mass of molecules. In
this derivation, we used the stochastic partial integration formula, see
Appendix. From Eq. (\ref{Forward}) and the Ito formula \cite{handbook}, we
obtain 
\begin{equation}
\tilde{D}D\mathbf{r}^{i}=\tilde{D}\mathbf{u}^{i}=\partial _{t}\mathbf{u}%
^{i}+\sum_{j}\tilde{\mathbf{u}}^{j}\partial _{j}\mathbf{u}^{i}-\sum_{jk}\nu
^{jk}\partial _{j}\partial _{k}\mathbf{u}^{i}.
\end{equation}%
Here the above expression corresponds for the general noise tensor $\mathbf{B%
}$. For the following discussion to derive the NS equation, we consider the
case of $\mathbf{B=}$ $\sqrt{2\nu }\mathbf{I,}$ where $\mathbf{I}$ is a unit
matrix.

\section{Entropy Variation}

Differently from the ideal case, $\hat{s}$ should be treated as a functional
of $\mathbf{r}$ and we need to specify $\delta _{\lambda }\hat{s}$ in terms
of Eq. (\ref{vari}). For this purpose, we employ a following simple model.
Suppose that the fluid-dynamical time scale $\tau _{hyd}\equiv \rho /\dot{%
\rho}$ is much larger than that of microscopic degrees of freedom, $\tau
_{mic}$, inside a fluid element. Then local thermal equilibrium should be
achieved, recovering the ideal fluid where $\hat{s}$ is constant $(\delta
_{\lambda }\hat{s}=0)$. Therefore, we expect that the entropy variation is
expressed in powers of $\tau _{mic}/\tau _{hyd}$ in such a way that the
ideal fluid case is recovered in the vanishing limit of $\tau _{mic}$. We
then write $\delta _{\lambda }\hat{s}=\delta _{\lambda }(a_{1}\tau _{mic}%
\dot{\rho}/\rho +a_{2}(\tau _{mic}\dot{\rho}/\rho )^{2}+\cdots )$, where $%
a_{i}%
{\acute{}}%
s$ are expansion coefficients. The lowest order truncation gives 
\begin{equation}
\delta _{\lambda }\hat{s}=\delta _{\lambda }\left( g(\rho )\dot{\rho}\right)
,  \label{delta_s}
\end{equation}%
where $g(\rho )$ is an arbitrary function of $\rho $. For the stochastic
variation, $\dot{\rho}$ is interpreted as $(D+\tilde{D})\rho /2$. Note that $%
\delta _{\lambda }\hat{s}$ is the virtual change of $\hat{s}$ associated
with the variations and does not necessarily satisfy the thermodynamic
principles such as $\delta _{\lambda }\hat{s}\geq 0$.

\section{Navier-Stokes equation}

By substituting Eq. (\ref{delta_s}) into Eq. (\ref{ls}), the condition of $%
\delta _{\lambda }I=0$ for the arbitrary function $\mathbf{f}(\mathbf{r},t)$
leads to 
\begin{equation}
\rho (\partial _{t}+\mathbf{v}_{m}\cdot \nabla )\mathbf{u}^{i}+\partial
_{i}(P-\mu \nabla \cdot \mathbf{v}_{m})-\sum_{j}\partial _{j}(\eta \partial
_{j}\mathbf{u}^{i})=0.  \label{general}
\end{equation}%
Here we used that $\mu =-\rho ^{3}g(\rho )/m(\partial T/\partial \rho )_{%
\hat{s}}$ and $\eta =\nu \rho $. The contribution from $\delta _{\lambda }%
\hat{s}$ effectively changes pressure by $\mu \nabla \cdot \mathbf{v}_{m}$.
The coefficient $\mu $ is known as the second coefficient of viscosity.

As was pointed out, the fluid velocity of the NS equation is not $\mathbf{u}$
but $\mathbf{v}_{m}$. Eliminating $\mathbf{u}$ using Eq. (\ref{mass_v}), Eq.
(\ref{general}) is finally re-expressed as 
\begin{eqnarray}
&&\rho (\partial _{t}+\mathbf{v}_{m}\cdot \nabla )\mathbf{v}%
_{m}^{i}+\sum_{j}\partial _{j}[(P-\zeta \nabla \cdot \mathbf{v}_{m})\delta
^{ij}-\eta e_{ij}^{m}]  \notag \\
&&-\sum_{j}\partial _{j}\left( \eta \partial _{j}\left( \frac{\eta }{\rho }%
\partial _{i}\ln \rho \right) \right) =0,  \label{NS}
\end{eqnarray}%
where 
\begin{equation}
e_{ij}^{m}=\partial _{j}\mathbf{v}_{m}^{i}+\partial _{i}\mathbf{v}_{m}^{j}-%
\frac{2}{3}(\nabla \cdot \mathbf{v}_{m})\delta _{ij}.
\end{equation}%
We thus identify $\eta $ as the shear viscosity and $\zeta =\mu +2\eta /3$
as the bulk viscosity. The last term is not only of second order for the
magnitude of fluctuations $\nu =\eta /\rho $, but also of third order for
the spatial derivative $\nabla $. In accordance with the approximation used
in the NS equation, this term should be discarded as a higher order
correction. In this sense, Eq. (\ref{NS}) is completely equivalent to the
compressible NS equation.

\section{Concluding Remarks}

In this letter, we showed for the first time that the stochastic variational
method can be used to derive the full Navier-Stokes equation. The basic
ingredients of this approach are the classical action of the ideal fluid and
the stochastic motions of fluid elements which are induced by the white
noise.

The present derivation shows clearly that there are two different physical
origins for the shear and bulk viscosities. The shear stress tensor and a
part of the bulk viscous pressure are obtained through the stochastic
motions of fluid elements, while the entropy variation in the potential term
affects only the bulk viscous pressure (the second coefficient of
viscosity). It is easy to see that the entropy variation due to the shear
stress tensor $e_{ij}^{m}$ gives only higher order corrections. Thus, the
shear viscosity in the NS equation is not altered by such a modification.

So far, we have emphasized the mathematical aspects of SVM and did not
discuss its physical background. To see it, note that the noise introduced in
Eq. (\ref{eqn:sde}) is related directly to the transport coefficients. For
example, we can show that $\eta$ satisfies the Einstein relation, 
\begin{equation}
\eta =\frac{\rho }{3}\int_{0}^{\infty }dtE[\delta \widehat{\mathbf{v}}%
(t)\cdot \delta \widehat{\mathbf{v}}(0)],  \label{fd2nd}
\end{equation}%
where $\delta \widehat{\mathbf{v}}=d\mathbf{r}/dt-\mathbf{u=}\sqrt{2\nu }d%
\mathbf{W/}dt$. This is nothing but the realization of the fluctuation-dissipation theorem 
and it appears as a natural consequence of SVM. We thus conclude that SVM 
possesses not only the well-defined mathematical structure but also a reasonable 
mechanism of dissipation. 

Above results show that the SVM approach is considered as a promising
framework, and can be extended to more general dissipative phenomena which
are not in the scope of the NS equation. In fact, generalization of a
diffusion equation is done in SVM \cite{kk_svm}. Another possible example
can be found in the soft matter physics. One formulation of such dissipative
equations is based on Onsager's variational method \cite{doi} so that the
comparison of SVM to Onsager's method will clarify the physics of variational approaches for the
dissipative phenomena.

As shown, the present result of SVM specifies the form of the higher order
correction to the NS equation. If SVM is a reliable approach, this higher
order term neglected in the NS equation should be considered seriously. The
structure of Eq.(\ref{general}) reminds that of the generalized
hydrodynamics proposed by Brenner \cite{brenner,other}, but in the SVM
approach, the difference of the two velocity fields appears as the higher
order correction to the NS equation. One example of the importance of higher
oder terms obtained by SVM plays in fact a crucial role in diffusion
processes. The application of SVM to a diffusion process leads to a
generalized form of the diffusion equation \cite{GenDif} which contains memory
effects. The existence of the higher order term guarantees Fick's law \cite{kk_svm}.

As mentioned above, in spite of very attractive features, the general
availability of SVM is not yet established. This question should be
investigated by the applications of SVM to field theoretical systems,
relativistic systems and complex fluids.

\hspace{1cm}

This work was financially supported by CNPq, FAPERJ, CAPES and PRONEX.

\appendix 

\section{stochastic partial integration formula}

The time variable is discretized as 
\begin{equation}
t_{j}=a+j\frac{b-a}{n},~~~j=0,1,2,\cdots ,n.
\end{equation}%
Then we can show 
\begin{eqnarray}
\lefteqn{\int_{a}^{b}dtE[\left\{ D\bf X(t)\right\} \cdot \bf Y(t)+\bf
X(t)\cdot \tilde{D}\bf Y(t)]}  \notag \\
&=&\lim_{n\rightarrow \infty }\sum_{j=0}^{n-1}E\left[ (\mathbf{X}_{j+1}-%
\mathbf{X}_{j})\frac{\mathbf{Y}_{j+1}+\mathbf{Y}_{j}}{2}\right] \frac{b-a}{n}
\notag \\
&&+\lim_{n\rightarrow \infty }\sum_{j=1}^{n}E\left[ \frac{\mathbf{X}_{j}+%
\mathbf{X}_{j-1}}{2}(\mathbf{Y}_{j}-\mathbf{Y}_{j-1})\right] \frac{b-a}{n} 
\notag \\
&=&\lim_{n\rightarrow \infty }\sum_{j=0}^{n-1}E[\mathbf{X}_{j+1}\mathbf{Y}%
_{j}-\mathbf{X}_{j}\mathbf{Y}_{j-1}]  \notag \\
&=&E[\mathbf{X}(b)\mathbf{Y}(b)-\mathbf{X}(a)\mathbf{Y}(a)].
\end{eqnarray}%
This is called the stochastic partial integration formula \cite{zm}.

\end{document}